# DESIGN PATTERNS FOR SELF ADAPTIVE SYSTEMS ENGINEERING

Yousef Abuseta and Khaled Swesi

Computer Science Department, Faculty of Science
Al-Jabal Al-Gharbi University, Libya

*ABSTRACT*

*Self adaptation has been proposed to overcome the complexity of today's software systems which results from the uncertainty issue. Aspects of uncertainty include changing systems goals, changing resource availability and dynamic operating conditions. Feedback control loops have been recognized as vital elements for engineering self-adaptive systems. However, despite their importance, there is still a lack of systematic way of the design of the interactions between the different components comprising one particular feedback control loop as well as the interactions between components from different control loops . Most existing approaches are either domain specific or too abstract to be useful. In addition, the issue of multiple control loops is often neglected and consequently self adaptive systems are often designed around a single loop. In this paper we propose a set of design patterns for modeling and designing self adaptive software systems based on MAPE-K. Control loop of IBM architecture blueprint which takes into account the multiple control loops issue. A case study is presented to illustrate the applicability of the proposed design patterns.*

## 1. INTRODUCTION

The development of today's software systems has become a very difficult task to accomplish. This is due to the dynamic and heterogeneous nature of these systems. It is rather difficult to build a software system that meets all requirements and make it survive without the need for a change either in response to new user requirements or to changing environmental conditions. Furthermore, many of these systems cannot afford too long downtime while the source code is being modified. Traditional top-down engineering approaches are insufficient to handle the complexity and evolution problems inherent in decentralized, continually evolving software [19]. Therefore, there is an increasing need for the software systems to be able to self adapt to accommodate new requirements not foreseen at design time or in response to changes on contextual conditions.

A self-Self-adaptive system (SAS) is a software-intensive system augmented with the ability to respond to a variety of changes that may take place in their environment, goals, or the system itself by adapting its structure and behavior at run-time autonomously [5, 22]. This vision requires shifting the human role from operational to strategic. Humans define high level policies that state how a system should react to changes and the system then carries out corrective changes (adaptations) autonomously at run-time [6].

Research in self-adaptive systems have been conducted within the different areas of software engineering, including requirements engineering [2], software architecture [9, 10, 25], middleware [18], and component-based development [23].

Feedback control loops [20] from the control theory area have been identified as vital elements in engineering self-adaptive software systems. Such control loops are typically organized by means of four components that are responsible for the fundamental functions of self-adaptation: Monitor, Analyze, Plan, and Execute, often referred to as the MAPE loop as in the IBM architecture blueprint [13]. In [7] these functions are referred to as collect, analyze, decide, and act.

Despite the importance of feedback loops regarding the introduction of self adaptability to today's software systems, there is still a lack of systematic way of the design of the interactions between the different components comprising one particular feedback control loop. Existing approaches are either too specific for some domains or highly abstract to be useful in modeling a wide range of domains. Also, little attention has been given to modeling and designing multiple interacting control loops since most approaches assume the existence of only one single control loop and design their self adaptive systems accordingly. In addition, the responsibilities of each component in the feedback control loop are not defined properly and consequently each SAS designer tends to define them differently. The monitor responsibility, for instance, is often defined in a way that overlaps with the analyzer's where the monitor accomplishes the data collection, aggregation and reasoning tasks. This contradicts with the separation of concerns design principle. Such a design principle contributes to a clean and flexible design of software systems in general and self adaptive systems in particular.

In this paper we introduce a set of design patterns which serves as an assisting tool for software designers of self adaptive systems. Such patterns support reuse of known solutions and evaluated ideas and architectures taken from well designed and accepted approaches for the area of self adaptive systems engineering. Our design patterns are based on the feedback control loop proposed by the IBM architecture blueprint [13].

The rest of the paper is organized as follows. Section 2 reviews some background issues related to our proposed design method of self adaptive systems. Section 3 introduces some related work on the self adaptive software systems development. Section 4 presents our proposed set of design patterns that depicts the well known solutions and ideas about the self adaptive systems engineering. In section 5, an evaluation case study is employed to evaluate and show the applicability of the proposed design patterns. The paper is concluded in section 6 with outlined directions for future work.

## 2. BACKGROUND

### 2.1 Feedback Control Loop

Feedback loops provide the generic mechanism for self-adaptation [3]. A feedback loop is a control loop where the output of the controlled system is fed back to the input. It allows therefore to adjust operations according to differences between the actual output and the desired output. In other words, feedback control loops are entities that observe the system and initiate adaption. A feedback loop typically involves four key activities: collect, analyze, decide, and act [7]. Sensors collect data from the running system and its environment which represents its current state. The collected data are then aggregated and saved for future reference to construct a model of past and current states. The data are then analyzed to infer trends and identify symptoms. The planning activity then takes place and attempts to predict the future and prepare change plan to act on the running system through a set of effectors or actuators [3].

### 2.2 Managed and Managing Systems

Self adaptation capabilities can be introduced to the software system either internally or externally [26]. In the internal approach, the adaptation logic (managing system) is intertwined with the core application (managed system) which may take the form of the exception handling. In this case, the adaptation engine is system dependent and thus difficult to maintain, evolve, and reuse. In contrast, in the external approach, the concerns of the adaptation logic are separated from the core application. Most of the existing approaches adopt the external approach since it enables the realization of some important software qualities such as the reusability and modifiability. The IBM architecture blueprint [13] is an example of this approach which is shown in Fig.1. As depicted in Fig.1, the managing system consists of four main activities: monitor, analyze, plan and execute. These activities share a knowledge base component which contains information about the system state as well as the policy engine that controls the system functioning. A set of sensors is used to collect the important data to the adaptation process and send them to the monitor for further processing while a set of effectors is used to apply the corrective changes stated in the plan.

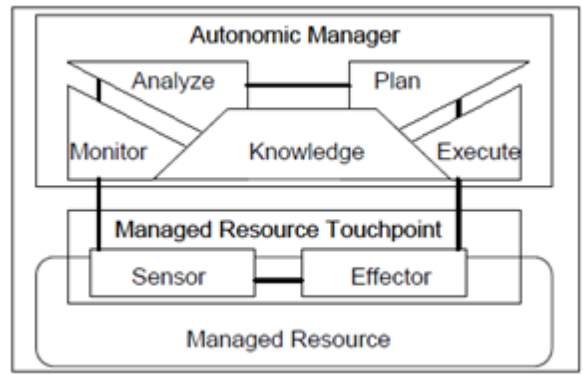

Fig. 1. Autonomic computing control loop [16]

## 2.3 Design Patterns

A Software design pattern is a repeatable solution to a usually occurring problem in software designs. It is a template for how to solve a problem that can be used in many different situations [8]. Design patterns [8,17] have fundamentally changed the way we approach the design of large software systems. Applying design patterns not only helps system designers take advantage of the software community's collective experience as captured in the patterns, it also enables others studying the system in question to gain a deeper understanding of how the system is structured, and why it behaves in particular ways.
In [8], design patterns have been classified into three categories as follows:

- Creational patterns which are concerned with creating the appropriate objects for a given case. This gives more flexibility to the program since choosing the right object to create is postponed to the runtime.
- Structural patterns which are concerned with composing groups of objects into larger structures.
- Behavioral patterns which define the communication between a set of objects in a software system and how the flow is controlled in such a system.

A number of design patterns belonging to the categories listed above plays a crucial role in designing self adaptive systems. The observer design pattern, for instance, can fit nicely when establishing the relationship between the monitor component and the sensor. Also, the Strategy design pattern can be used to design and implement the plan component which contains a set of possible actions executed when some particular conditions hold. Sometimes it is desirable and beneficial for the self adaptive systems to offer the option of reverting to previous system states (prior to the application of the plan changes). In the latter, the command design pattern comes in play and offers this feature. It is similar to the system restore option offered by Windows where the system might encounter some problems and choosing a restore point might solve the issue.

## 3. RELATED WORK

Several approaches have been proposed to address the design of self adaptive software systems. These approaches can be classified to requirements engineering , software architecture , middleware, and component-based development . Since our approach is related to the design patterns and software architecture in general, this section is dedicated to previous works conducted in these two specific areas. In [9, 11], Garlan et al proposed the Rainbow framework which provides general, supporting mechanisms for self-adaptation which can be customized for different classes of systems. It provides a language, called Stitch, to represent the adaptation knowledge using high-level adaptation concepts of strategies, tactics, and operators. The implementation of Rainbow is based on one large control loop that is in charge of all activities related to the self adaptability issue for the whole system under consideration. Ramirez and Cheng [24] describe a set of design patterns at the software design level to facilitate the construction of a self-adaptive software system. Gomaa et al [12] proposed several patterns for dynamically reconfiguring specific types of software architectures at run time [11]. In particular, they extended the concepts of dynamic change management introduced by Kramer and Magee [18] by introducing four design patterns to specify the behavior required to dynamically reconfigure master/slave, centralized, server/client, and decentralized architectures. Iftikhar and Weyns [14] propose ActiveFORMS, short for Active Formal Models, which contributes to the self adaptive systems with an approach that guarantees the verified adaptation behavior at design time and provides support for dealing with dynamic goals at runtime. Ben Said et al [1] proposed a set of design patterns targeting the real time systems. Iglesia [15] in his PhD thesis proposed and modeled the semantics for a set of MAPE-K formal templates for the design of self adaptive systems. In [27], a set of design patterns is introduced by the authors to model the various interactions between MAPE loops in decentralized control of self adaptive software systems.

# 4. PROPOSED DESIGN PATTERNS

Our proposed design patterns are built on some previous works carried out for addressing the development of self adaptive software systems. In particular, we use the concepts proposed by IBM architecture blueprint [13] for modeling the feedback control loops, namely the Monitor, Analyze, Plan, and Execute. We also take advantage of the work presented in [3, 6] in which the design space of self adaptive systems is presented as a set of dimensions whereas each dimension is defined by a design question. We also studied the design patterns introduced in [27] to model the various interactions between multiple MAPE loops of self adaptive software systems.

## 4.1 Metamodel of self adaptive systems

This section is dedicated to introduce the metamodel for the MAPE-K based self adaptive systems. It shows, in a high level view, the concepts involved in the MAPE-K control loop as well as the relationships and interactions between these concepts. This metamodel is an attempt to accommodate the various possible interactions between MAPE-K components which might emerge in many different situations or domains. In some situations, for example, a number of sensors must notify a monitor of some specific system properties where in others one sensor is linked to only one monitor. Likewise, it is sometimes necessary for the analyzer to accomplish its task is to be notified by a set of monitors each responsible for monitoring one particular part of the whole system. In addition, it might be necessary in some situations to have interactions between two components of the same type (monitor-to-monitor, analyzer-to-analyzer, etc) and therefore it is important to accommodate this interaction as well. Also, the notification process, from the sensor to the monitor for instance, can be conducted either periodically or based on some specific events. Regarding the adaptation application, a plan component might send the change plan, called strategy in the metamodel, to a number of execute components and similarly, an execute component could make use of a set of actuators to apply the change actions. Fig. 3 shows this metamodel using UML class diagrams. As it can be noticed from Fig. 3 the closed and continuous control loop starts with the measurement conducted by the sensors and ends by changes made through the actuators. The dotted arrow shows the process of the applying the changes plan which should cause the runtime system state to be brought back to a desirable and acceptable state according to some predefined goals and requirements.

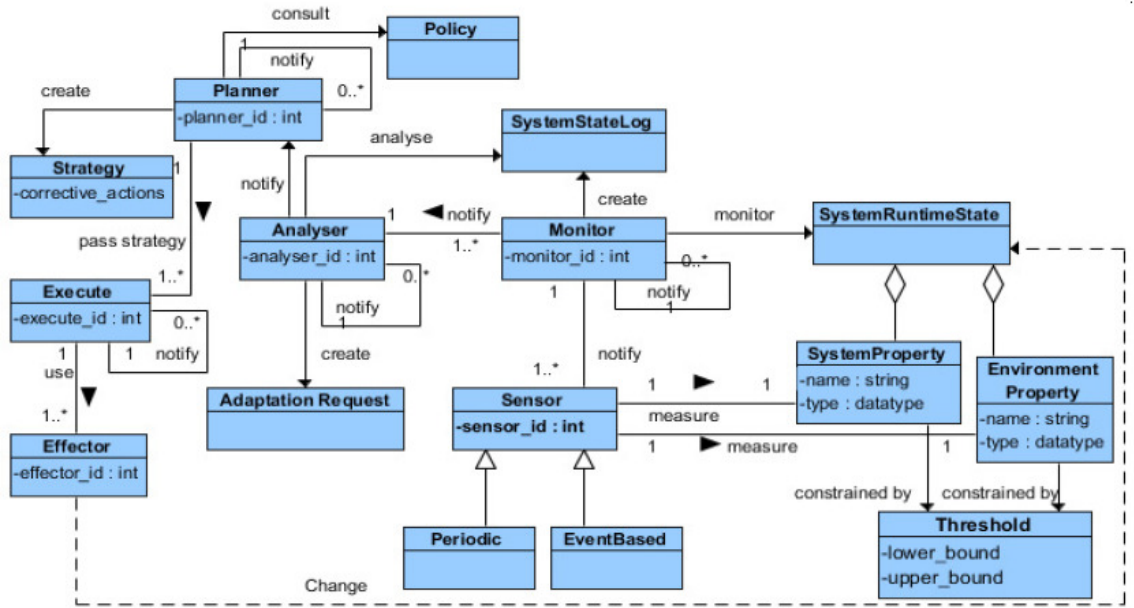

Fig. 2. Metamodel of MAPE-K based Self Adaptive Systems

## 4.2 SAS Design Patterns

Below, we introduce the five design patterns: SAS Monitor, SAS Analyzer, SAS Plan, SAS Execute and SAS Knowledge. Each design pattern proposed here is partially complied with the Gamma [8] template structure. Each pattern is described by six fields, namely the pattern name, intent, context, motivation, structural view and participants. UML diagrams are used here to depict the structural and behavioural views of the design patterns.

SAS Monitor design pattern

*Pattern Name*: **SAS Monitor**.

*Intent*: to establish the relationships between the components participating in accomplishing the monitoring activity of self adaptive systems using the MAPE control loop.

*Context*: this pattern is used in the first stage of the feedback control loop process aiming at introducing self adaptability to software systems. In this stage, *what properties to monitor question is answered*.

*Motivation*: the detection of a property threshold violation represents the trigger of the adaptation process in self adaptive systems. Therefore, in order to accomplish the detection process, a monitor component must be introduced.

*Structural view*: the components involved in the monitoring activity as well as their relationships are described using the UML class diagram shown in Fig. 3. As stated earlier, the relationship between the monitor and sensor is established using the observer design pattern. The monitor is interested in collecting data (property readings) from a set of sensors therefore it takes on the observer role while the sensors play the subject role. However, in this case the monitor (the observer) registers its interest of change notifications with a number of sensors (subjects) in a

one- to- many relationship. There is a concurrency taking place here since the sensors keep running simultaneously and notifying the monitor of any change occurred to the properties of their direct responsibility. To handle concurrency, most modern programming languages such as Java offer the multithreading technique where a number of threads are executing apparently at the same time. Therefore, the sensor class in Fig.3 is stereotyped with the word "Thread". Also since this concurrency leads to many threads attempting to access the same program (the monitor object), a synchronization mechanism should be put in place. This is realized in Java by preceding the operation in question (the update operation of monitor class) by the keyword *synchronized*.

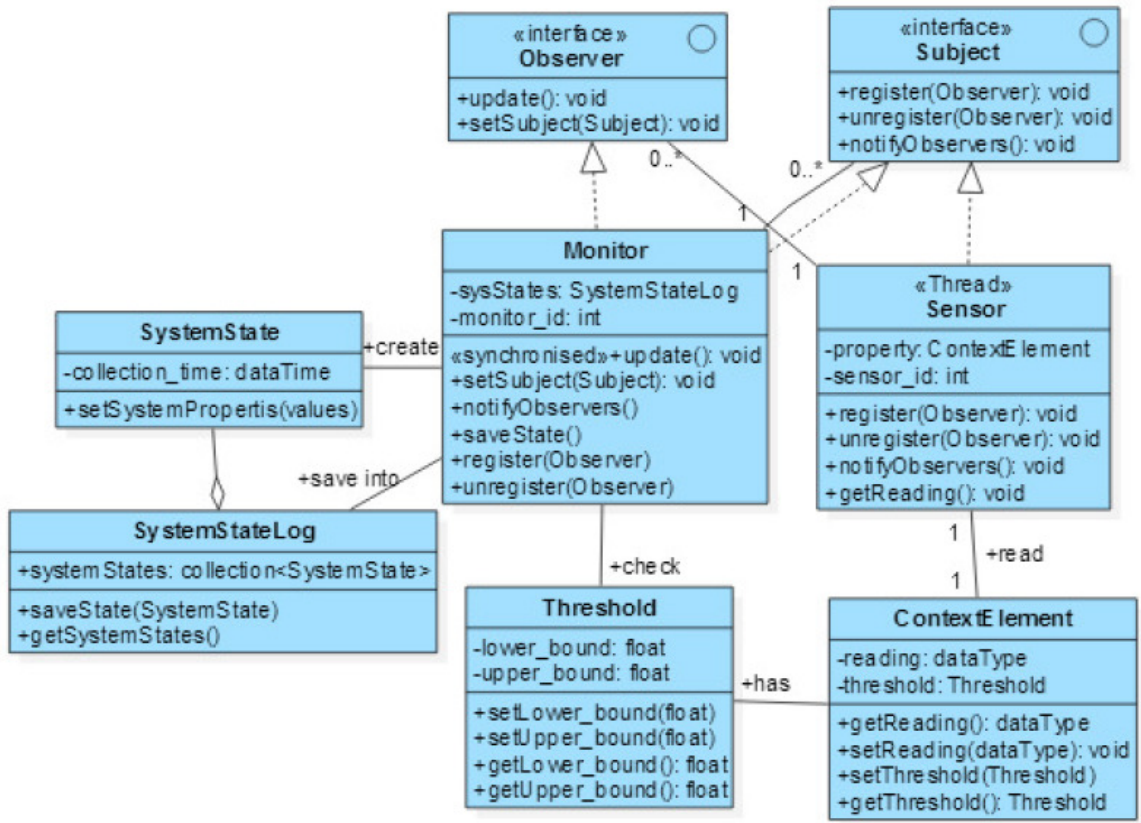

Fig. 3. UML Class diagram for the monitoring activity components.

*Participants*: lists the classes involved in the pattern and describes each class's responsibilities.

**Sensor.** Its sole responsibility is to collect data about the system property of high interest to the adaptation process and then send it to the monitor. Accomplishing this task can be conducted at fixed delay, in response to an event and/or on demand. When adopting the fixed delay form, the sensor sends the system property measurement to the monitor each x seconds where x > 0. However, the sensor may send the property measurement once the event of property threshold violation has occurred not waiting for the current time window to finish. Therefore, there are two kinds of sensor namely the time-triggered and event-triggered sensors.

**System property.** Also referred to as the context element, this is the property that is of a direct connection and great interest to the adaptation process. This property is the target of the monitoring activity and the main concern of the monitor component is to keep its value within a desirable or acceptable range. Often, a threshold is used to accomplish this task. Examples of system properties include server load, server throughput, response time and bandwidth usage. The system property contributes to the runtime system state.

**Environment property.** Tthe environment is defined as any external actor that affects the system in some way. Therefore, the environment property represents any contextual information that is external to the system in question and contributes to its runtime state. Examples of such properties include the time of operating, the current client connections in client-server architecture, etc.

**Threshold.** this is the value that the monitor component will compare against to decide whether the current value of the system property is still within a desirable or acceptable range. A threshold might have two values for the lower and upper bounds. An example of a threshold would be if server CPU load becomes greater than 50%, or if load changes by more than 20%.

**System runtime state.** At runtime, the system state is represented by the combination of the values of system properties and the properties representing the environment or the context within which the system is operating. Each system has a desirable state driven by its goals and non functional requirements. Often the deviation from this desirable state is the trigger of the adaptation process.

**Monitor.** It is responsible for collecting data from a set of sensors, filtering and aggregating this data and sending it to the analyzer component for any possible symptoms of system goals violation. The combined data collected from the sensors represent and form the system (or subsystem) runtime state. Conceptually, the monitor plays the role of the master in the master-slave relationship while the sensors play the slave roles. However, in terms of software design patterns, the observer pattern is used here where the monitor plays the observer role while the subject role is played by the sensor(s).

*Behavioral view.* The behavioral view depicts the interactions between the different components involved in the monitoring activity using UML Sequence diagram as shown in Fig. 4.

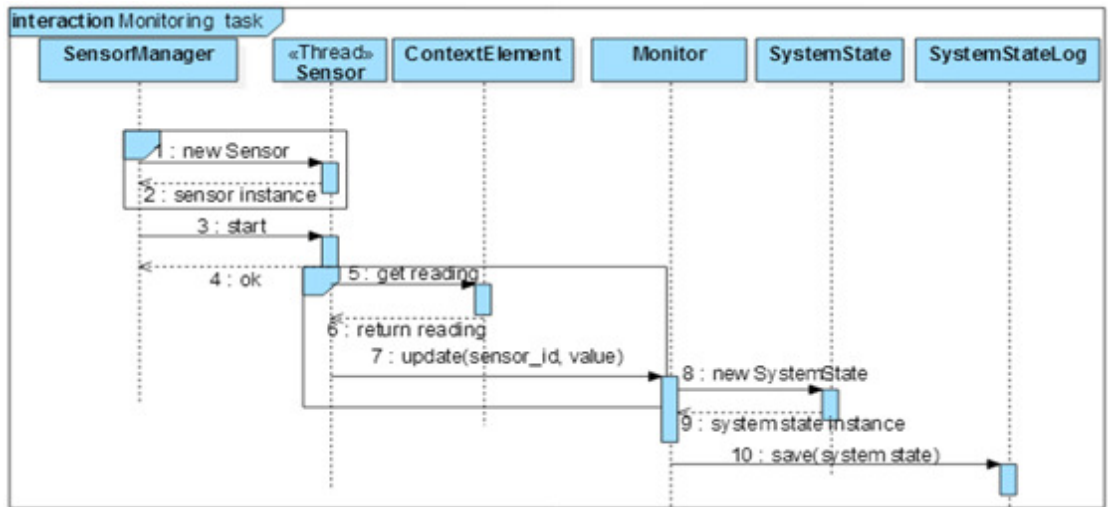

Fig. 4. The behavioral view of the monitoring activity of MAPE-K

SAS Analyzer design pattern

*Pattern Name*: **SAS Analyzer**.

*Intent*: to establish the relationships between the components participating in accomplishing the analyzer activity of self adaptive systems using the MAPE control loop.

*Context*: this pattern is used in the stage of evaluating the collected data by the monitor component for any possible symptoms of system goals and requirements violation. In this stage, the question of whether an adaptation is required or not is answered.

*Motivation*: detection of a system property (or system state violation) threshold violation represents the trigger of the adaptation process in self adaptive systems. Therefore, in order to accomplish the detection process, an analysis activity must be carried out which is performed here by the SAS analyzer pattern.

*Structural view:* the components involved in the analyzer activity of self adaptive systems as well as their relationships are described using the UML class diagram shown in Fig. 5. The central class of this activity is the analyzer which contains the update operation where it receives the collected data (SystemStateLog) from the monitor. It also notifies the plan component of any required adaptation. Therefore, it is linked with the monitor and plan components using the Observer design pattern where it plays both the subject role (with the plan) and the observer role (with the monitor) and thus has to implement two interfaces, namely the observer and the subject. The symptom class is mainly composed of an associative array where a key-value entry is used to store each symptom. The key can represent the event of the symptom (e.g. database connection failure) and the value is representing the conditions associated with this event. Different implementations of different programming languages offer classes and interfaces to implement this which include the Map (Java), dictionary (Python) and associative arrays (PHP).

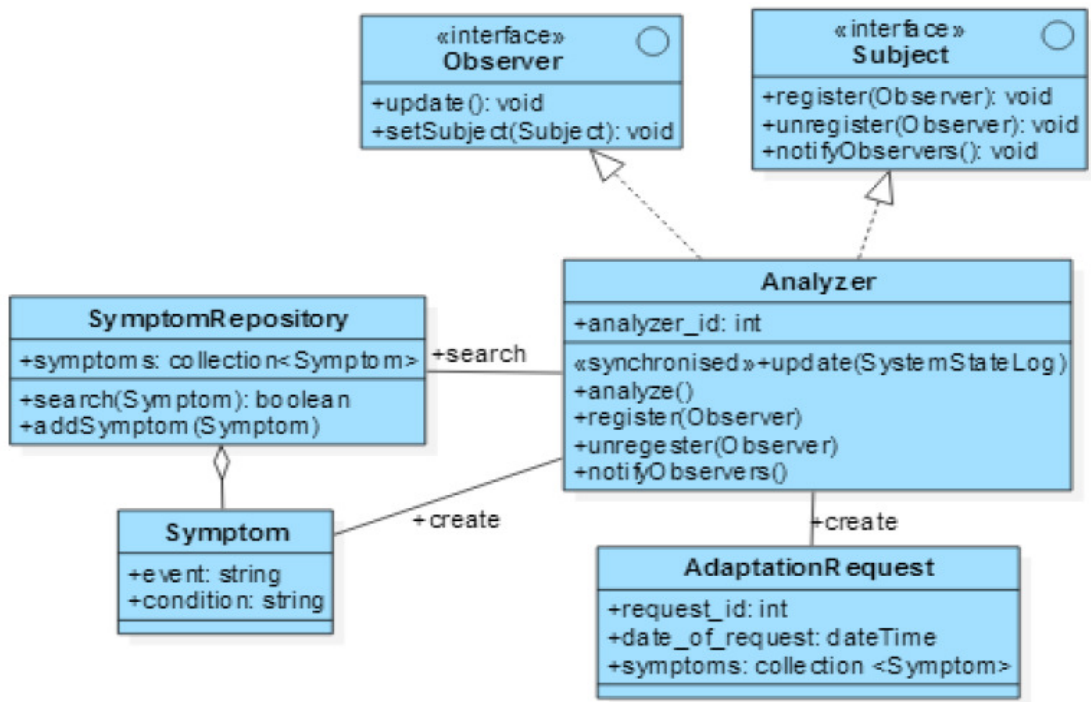

Fig. 5. UML Class diagram for the Analysis activity components

*Participants:* lists the classes involved in the pattern and describes each class's responsibilities.

**Analyzer.** Its responsibility is to receive collected and logged data (from monitor) representing the system (or subsystem) state history and analyze them for any possible symptoms of system goals and requirements violation. The analyzer notifies the plan of any necessary adaptations via sending an adaptation request.

**Symptom.** Represents one of the undesirable states that the system in question must detect and take corrective actions against. A highly loaded server is an example of such symptoms. Symptoms work with a set of combined conditions and when these conditions are satisfied, the analyzer raises an adaptation request signal and sends it, along with the necessary information, to the plan component.

**AdaptationRequest.** Once the analyzer analyses the received data from the monitor and decides that some symptoms exist, an adaptation request is created and sent to the plan component along with the necessary information. The latter include the event describing the symptom (e.g. highly loaded server) and the occurrence of this event in a specified time window (e.g. last two hours).

**SymptomRepository.** It contains a set of predefined symptoms that the system in question should avoid and heal up from. It also provides a facility to add new emerging symptom at runtime via the addSymptom operation. This component is usually part of the knowledge base of the feedback control loop.

*Behavioral view:* The behavioral view depicts the interactions between the different components involved in the analysis activity using UML Sequence diagram as shown in Fig. 6. Notice that both the analyzer and plan components have the update operation, which is part of the Observer interface, since they both play the observer role in this interaction. The analyzer receives the collected data from the monitor (the trigger of the analysis activity) and the plan receives the adaptation request signal from the analyzer once a symptom of undesirable system state is detected.

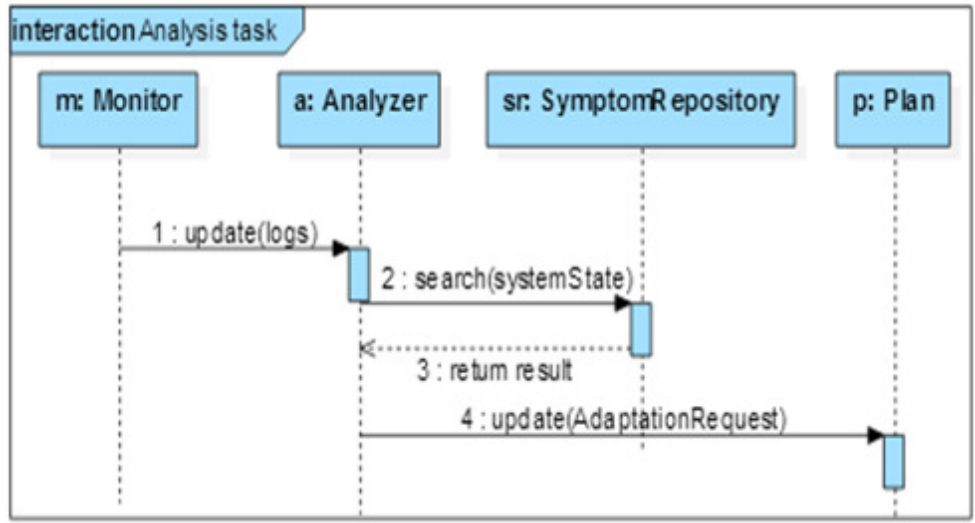

Fig. 6. The behavioral view of analysis activity of MAPE-K

SAS Planner design pattern

*Pattern Name*: **SAS Plan**.

*Intent*: to establish the relationships between the components participating in accomplishing the plan activity of self adaptive systems using the MAPE-K control loop.

*Context*: this pattern is used in the stage of constructing the change plan which is composed of a set of corrective actions in response to an adaptation signal raised by the analyzer component. In this stage, *the questions of what actions to be taken and in what order.*

*Motivation*: When modeling a self-adaptive systems, designers need to specify the adaptation strategy to use to calculate the adaptation decisions.

*Structural view*: the components, and their relationships, involved in the plan activity of self adaptive systems are described using UML class diagrams as shown in Fig.7.

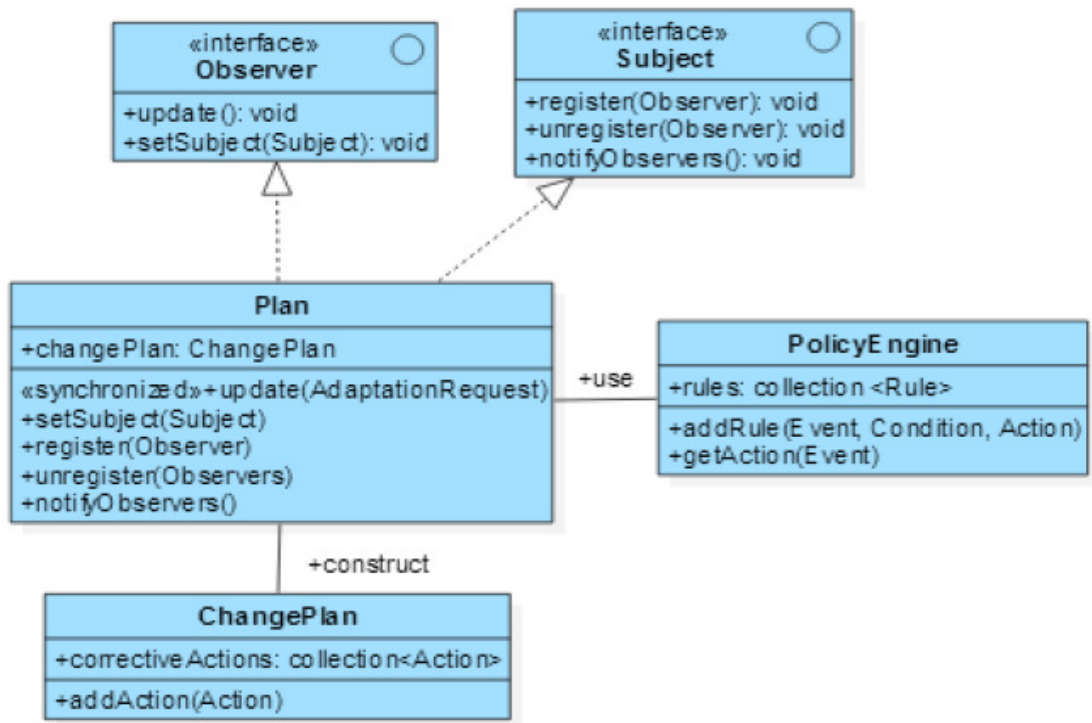

Fig. 7. UML Class diagram for the Planning activity components

*Participants:* lists the classes involved in the pattern and describes each class's responsibilities as follows:

**Plan.** It is responsible for constructing the change plan in response to an adaptation request received from the analyzer. The plan component uses the policy engine for accomplishing its task and then sends the constructed change plan to the execute component to dispatch these changes. The plan is linked with the analyzer and execute using the observer design pattern where it takes on both the observer and subject roles respectively and thus implements the Observer and Subject interfaces.

**PolicyEngine.** It contains the policies (high level goals) that control the operating and functioning of the system in question. Policies take the form of Event-Condition-Action (ECA) rules which determine the actions to be taken when an event is raised provided some specific conditions are met. A general form of a policy rule can be written as:

on *event* if *condition* do *action.*

The policy engine belongs to the knowledge base of the feedback control loop. It provides the necessary interface for the system administrators to define and modify the policies of the system at hand.

**ChangePlan.** It contains the actions that should be dispatched to the execute component in order to perform the adaptation and corrective actions. It is often called the strategy in which the actions are performed in specific and logical order.

*Behavioral view:* The behavioral view depicts the interactions between the different components involved in the planning activity using UML Sequence diagram as shown in Fig.8.

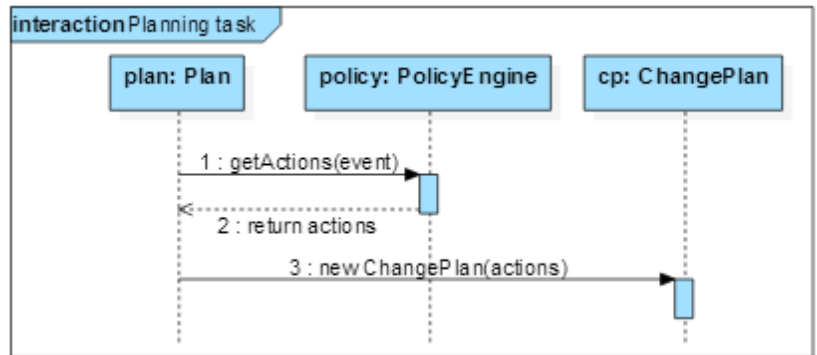

Fig. 8.The behavioral view of planning activity of MAPE-K

SAS Execute design pattern

*Pattern Name*: **SAS Execute**.
*Intent:* to establish the relationships between the components participating in accomplishing the execute activity of self adaptive systems using the MAPE control loop.

*Context:* this pattern is used in the stage of executing the adaptation actions or change plan that is received from the plan component. These actions must be executed in some specific order (sequentially or concurrently or maybe mixed of the two) as stated in the plan. The execute component uses a set of actuators or effectors to apply the required changes to the system and usually involve setting new values to the system properties which are collectively constitute the system state. In this stage, *what properties to change question is answered.*

*Motivation*: this pattern models and design the last activity of the MAPE-K which involves dispatching some corrective actions to a number of effectors in order to bring the system back to a desirable or acceptable state. Executing the adaptation in the order stated in the change plan is crucial for a successful transition from an undesirable to desirable state.

*Structural view*: the components and their relationships are described using UML class diagrams as shown in Fig.9. The central class of this activity is the executor which contains the update operation where it receives the change plan (corrective actions) from the plan. Once it has received the corrective actions, it dispatch them to a set of effectors to apply the changes to the target system and environment properties (referred to as context element in Fig. 9). Therefore, it is linked with the plan and effector components using the Observer design pattern where it plays both the observer role ( with the plan) and the subject role (with the effector) and thus has to implement two interfaces, namely the observer and the subject.

*Participants:* lists the classes involved in the pattern and describes each class's responsibilities.
**Executor.** It is responsible for sending the corrective actions to one or more effectors in a specific order.

**Effector.** It is responsible for applying changes to system or environment properties according to some actions received from the executor component.

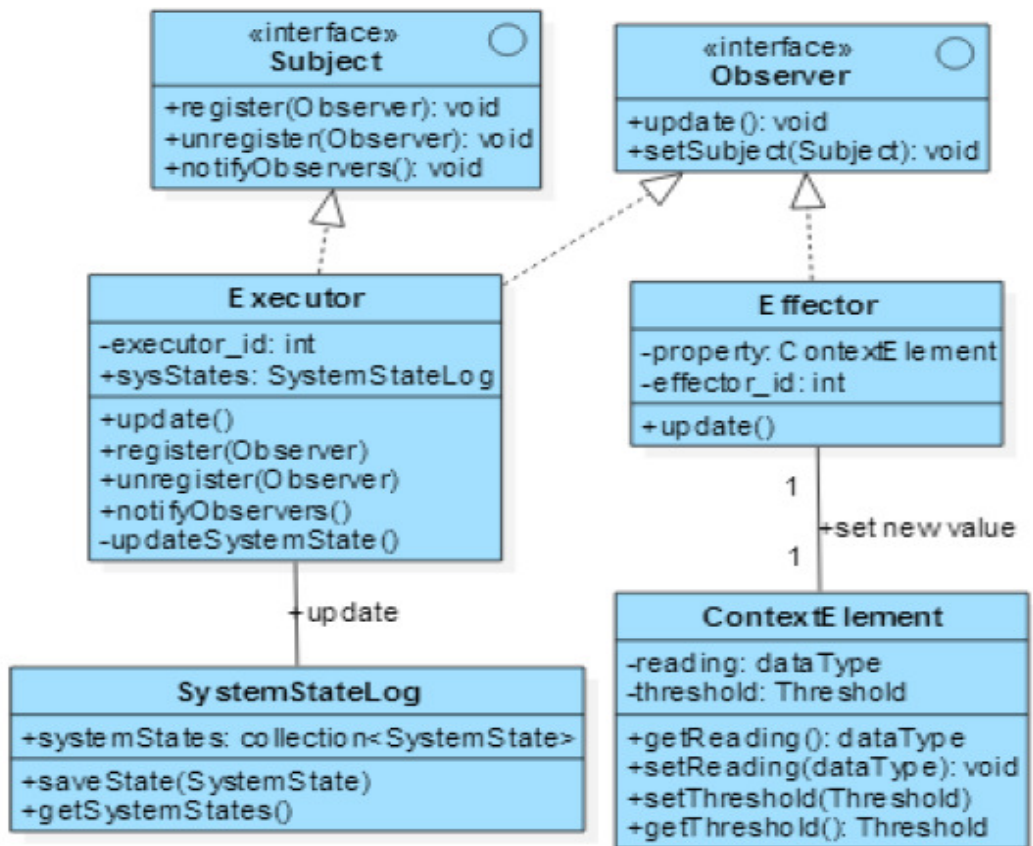

Fig. 9. UML Class diagram for the Execute activity components

*Behavioral view:* the behavioral view depicts the interactions between the different components involved in the execute activity using UML Sequence diagram as shown in Fig.10.

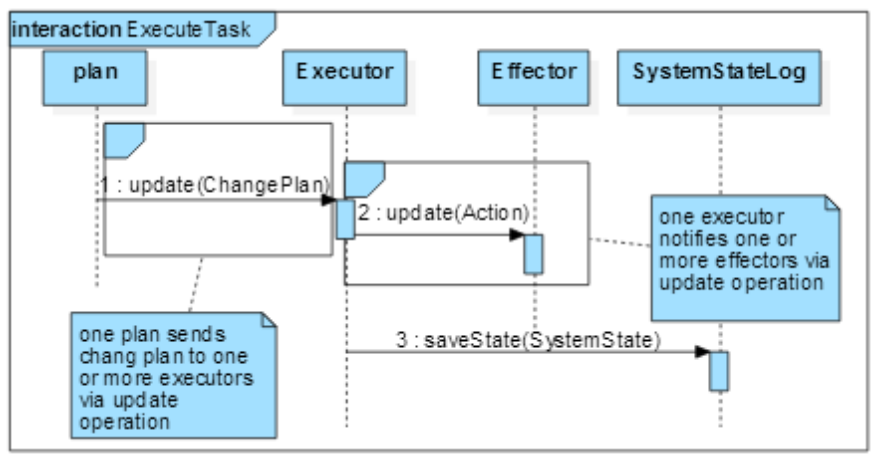

Fig. 10. The behavioral view of Execute activity of MAPE-K

## 5. EVALUATION CASE STUDY

We evaluate the effectiveness of our devised design patterns in introducing self adaptation to an existing systems in two aspects: (1) How easy to introduce the self adaptation capability to a software system? (2) How flexible are the design patterns in accommodating the different scenarios of feedback control loop interactions?. To carry out the evaluation , we introduce here the case study of augmenting Virtual Learning Environments (VLE), such as Moodle and Blackboard, with self adaptive capabilities. Architecturally, these environments are viewed and designed using the client-server style where such environments provide E-learning services through a set of servers (server farm) which can be accessed by a number of learners as well as instructors (clients). One of the major concern of these environments is to achieve scalability (scale up/down) to either accommodate more clients accessing the system or to cut down the operating cost by switching off some servers.  When managing a client-server based software system, the load balancing technique largely contributes to solving many server related problems including the response time, throughput and scalability.  Therefore, the feedback control loop is necessarily composed of components concerned with the load balancing mechanism. Such components include load monitor, load analyzer,  load balancing planner and load balancing executor.  A knowledge base component is used here by these components to achieve their tasks. The candidate properties that can be identified here for monitoring along with the components that they may apply to are as follows:

- load  and throughput properties which are associated with the server component. Load expresses how many processes are waiting in the queue to access the server processor or can be measured by the CPU usage while throughput measures the amount of messages that a server processes during a specific time interval.
- latency property is associated with the client component  since latency is measured from the client machine which is the time taken from sending a request  to receiving a reply.
- cost property  is associated with the Server component (number of active servers).

The adaptation process is triggered by the outcome of the observation of the server load. The possible adaptations, which depend on the current system state and on the system goals, are as follows:

- *Add server*: this operation causes the addition of a new server to the server farm as a corrective action to the sever high load provided that this wouldn't affect the operating cost.
- *Remove server*: this operation causes the removal of a server from the server farm either because of a low load at specific time or as a response to an unresponsive server.

As stated earlier, the main objectives of this system is to keep the E-Learning services available to learners and instructors while maintaining the operating cost at a specific range. Thus, the availability and operating cost issues are of a great interest to the feedback control loop. The process of ensuring the latter Quality of Service (QoS) attributes can be described as follows: There will be a number of sensors, one for each active server to measure the current reading of the server load and send it to the monitor. Collectively, these readings represent the runtime system state of the server farm. Each sensor reports its reading of server load once the latter has changed by 20% for instance (change of load reading is affected by the access requests of the learners and instructors). Upon receiving a new reading from any sensor, the monitor constructs a new runtime system state and save it in the system log and send it to the analyzer. Then the analyzer analyzes the received data for any possible symptoms of undesirable events. Three symptoms or events can be identified here, namely the *high load, very low load and* *unresponsive*. The *high load* event is raised when a server load becomes greater than 70% (upper bound threshold) while the very low load event is detected when the load is less than 5% (lower bound threshold) for some specific period (last two hours for instance). The *unresponsive* event is raised when no new reading of server load is received from the sensor for some specific period. The plan then constructs the corrective actions based on the system policies.

The two main policies of this system are:(1) add new server when a *high load event* is raised provided that the new number of active servers does not exceed the defined threshold (5 for example) , and (2) remove server from the system when either the *very low load* or *unresponsive* event is raised.

So far, we have demonstrated how to map the concepts presented in our design patterns to concepts in a real world example. To illustrate the flexibility of accommodating the different scenarios of feedback control loop interactions, we construct the possible organisation (shown in Fig. 11) of the self adaptive system which is tailored for this case study. A number of interactions can be recognized in Fig. 11. A monitor to monitor and execute to execute are two examples of an interaction where the two components are of the same type. We were able to model this kind of the interactions because we have designed the monitor and execute components (the same applies to the analyzer and plan) as both an observer and subject according to the Observer design pattern. In the *register* operation of the Subject interface, for example, the required parameter is of Observer type, therefore a monitor can register with another monitor to be notified of possible changes. Here, each monitor is responsible for collecting the server load of its subsystem (e.g. VLE.server_1 ) from the sensor and then report it to the monitor in the main control loop whose responsibility is to aggregate the collected data from the monitors and send it to the analyzer. Scalability (up/down) can be easily accomplished here since whenever we add/remove server to or from the server farm, we register or unregister the main monitor with or from the monitor of the recently added or removed server.

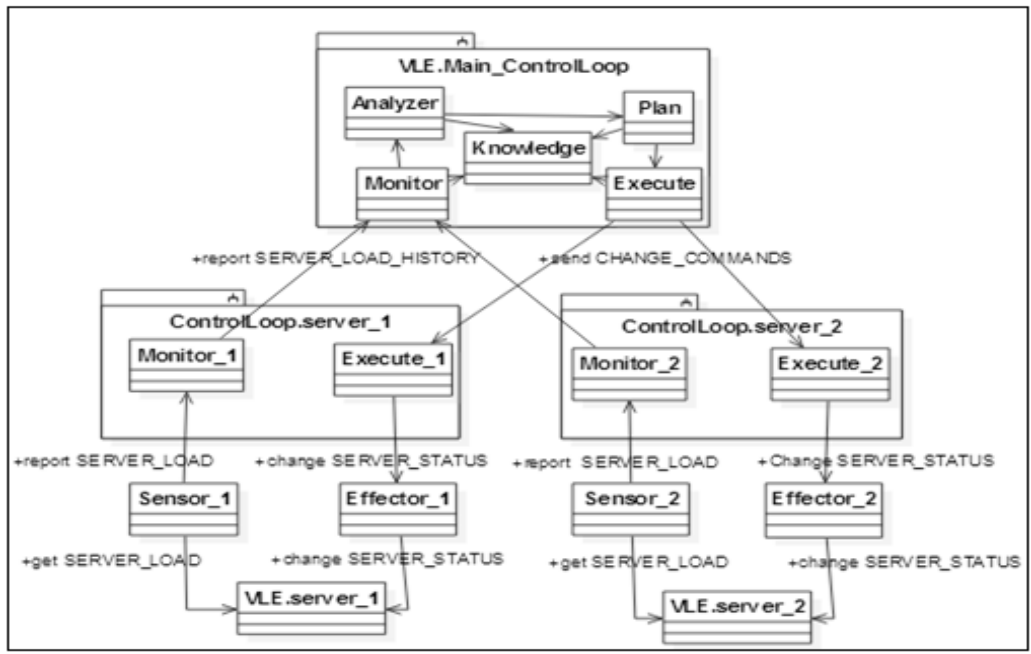

Fig. 11. Possible feedback control loop interactions for Self Adaptive VLE

# 6. CONCLUSIONS

In this paper, we have introduced a set of design patterns to model the four main activities of the MAPE-K control loop which are used to engineer self-adaptive systems (SAS). Accordingly, such patterns are termed as SAS Monitor, SAS Analyzer, SAS Plan and SAS Execute. Also, the interactions between the different MAPE-K components are modeled and designed. Such interactions include Sensor-Monitor, Monitor-Analyzer, Analyzer-Plan, Plan-Execute and Execute-Effector. We have applied the Observer design pattern for each interaction where all components (except for the sensor and effector) play both the observer and subject roles and thus implement two interfaces namely the observer and subject. This, besides the stated above interactions, enables the establishment of the interaction between two component of the same type (monitor for instance) which is needed in some situations. The E-learning system case study was used to illustrate the applicability of part of these design patterns. As future work and suggested research, the following issues need to be addressed:

- A more detailed case study to illustrate the application of our proposed deign patterns.
- The application of our design patterns to introduce the self adaptation capability to VLEs at the core services level. Such services include a self adaptive user interface and self adaptive learning paths. In the case study presented here, we were interested in providing the scalability at the hardware level (server farm). The server farm represents the platform where theVLEs are executed.
- Modeling and programming the different states that each MAPE component goes through during the management of the managed system.
- A complete lifecycle for the development of self adaptive systems.
- The impact of the architectural style of the system under development on the feedback control loops organization (decentralized for instance).

## AUTHORS

**Yousef Abuseta** received an M.Sc in Computer Engineering in 2003 from University of Hertfordshire, UK. In 2009, he received a PhD degree in Software Engineering (Autonomic systems) from Liverpool John Moores University, UK. He has been lecturing at Al-Jabal Al-Gharbi University since 2011 where he is working as a Senior Lecturer. His research interests include self adaptive systems, autonomic systems, design patterns, E-learning, Model Driven Development (MDD) and Service Oriented Architecture (SOA).

**Khaled Swesi** received an M.Sc in Computer Science in 2003 from Newcastle University, UK. In 2010, he received a PhD degree in Software Engineering from De Montfort University, UK. He joined Al-Jabal Al-Gharbi University in 2012 where he is working as a Senior Lecturer. His research interests include E-learning, AI and web-based applications.